\documentclass[aps,prl,preprint,showpacs]{revtex4}
\usepackage{graphicx, url, amssymb}
\begin{document}
\title{First Test of Lorentz Invariance in the\\ Weak Decay of Polarized Nuclei}

\author{S.E.~M\"uller}
\altaffiliation{Present address: Institute of Radiation Physics, Helmholtz-Zentrum Dresden-Rossendorf, Germany.}
\author{E.A.~Dijck}
\author{H.~Bekker}
\altaffiliation{Present address: Max Planck Institute for Nuclear Physics, Heidelberg, Germany.}
\author{J.E.~van~den~Berg}
\author{O.~B\"oll}
\author{S.~Hoekstra}
\author{K.~Jungmann}
\author{C. Meinema}
\author{J.P.~Noordmans}
\author{M. Nu\~nez Portela}
\author{C.J.G.~Onderwater}
\author{C.~Pijpker}
\author{A.P.P.~van der Poel}
\author{B.~Santra}
\author{A.~Sytema}
\author{R.G.E.~Timmermans}
\author{O.O.~Versolato}
\altaffiliation{Present address: Max Planck Institute for Nuclear Physics, Heidelberg, Germany.}
\author{L.~Willmann}
\author{H.W.~Wilschut}
\email{wilschut@kvi.nl}
\author{K.~Yai}
\altaffiliation{University of Osaka, Osaka, Japan}
\affiliation{Kernfysisch Versneller Instituut, University of Groningen,
  Zernikelaan 25, 9747 AA Groningen, The Netherlands}
\date{July 31, 2013}

\begin{abstract}
A new test of Lorentz invariance in the weak interactions has been
made by searching for variations in the decay rate of
spin-polarized ${}^{20}$Na nuclei. This test is unique to
Gamow-Teller transitions, as was shown in the framework of a
recently developed theory that assumes a Lorentz symmetry breaking
background field of tensor nature.
The nuclear spins were polarized in the up and down direction,
putting a limit on the amplitude of sidereal variations of the
form $|(\Gamma_{\mathrm{up}} -
\Gamma_{\mathrm{down}})|/(\Gamma_{\mathrm{up}} +
\Gamma_{\mathrm{down}}) < 3\times 10^{-3}$.
This measurement shows a possible route toward a more detailed testing of Lorentz symmetry in weak interactions.
\end{abstract}

 \maketitle
Lorentz invariance means that physical laws are independent of
boosts and rotations. It is at the basis of all known
interactions.
In the weak sector relatively few tests of Lorentz invariance have been made, even though
the understanding of the weak interactions has been crucial in
developing the standard model. In this work we consider a new test
that exploits the spin degrees of freedom in
$\beta$ decay, searching for a dependence of the nuclear lifetime on the orientation of the nucleus.
Recent theoretical work \cite{JN2013} enables
relating the present test to other possible Lorentz
symmetry tests in the weak interactions and put them in the overall
framework developed by Kosteleck\'y and coworkers~\cite{Colladay:1998fq}.
The CPT tests of neutral-meson \cite{mesonoscillations} or
neutrino \cite{Diaz1} oscillations and the $\beta$-decay endpoint
spectrum \cite{Diaz2} are also in the weak domain, however, they differ in nature.

We write the relative variation in the $\beta$-decay rate $\Gamma$ as
\begin{equation}\label{LIVbeta}
    \frac{d\Gamma}{\Gamma_0}= 1 + \vec{\beta}\cdot\left[ A\frac{\langle\vec{I}\rangle}{I}+ \xi_1 \hat{n}_1 \right]
    + \xi_2 \frac{\langle\vec{I}\rangle}{I}\cdot\hat{n}_2 .
\end{equation}
Here, $\Gamma_0$ is the standard model decay rate, with
$\vec{\beta}$ the velocity vector of the $\beta$ particle in units of the speed of light.
 The nuclear polarization of the parent nucleus is ${\langle\vec{I}\rangle}/{I}$.
$A$ is the $\beta$-asymmetry parameter in the standard model that violates
parity. Other parameters in the $\beta$ decay of spin-polarized
nuclei \cite{JN2013} are not relevant for this work.

Lorentz invariance violation (LIV) appears in
Eq.~(\ref{LIVbeta}) with magnitudes $\xi_1$, $\xi_2$ and directions
$\hat{n}_1$, $\hat{n}_2$ relative to the emission
direction of the $\beta$ particle and the polarization of the parent nucleus,
respectively. The directions $\hat{n}_i$ need not be identical. In
the theory of Ref.~\cite{JN2013} $\xi_i$ and $\hat{n}_i$ depend on
the nature of the transition, e.g.\ an allowed Fermi or Gamow-Teller
transition, or whether a transition is forbidden \cite{Jacob}.
 For the current discussion Eq.~(\ref{LIVbeta}) suffices.
An overview of current experimental and theoretical work is given
in Ref.~\cite{HWhimself}.

Early experimental work looked for a limit on the $\beta$-emission
anisotropy of unpolarized nuclei in forbidden $\beta$ decays of
$^{90}$Y, $^{137}$Cs, and $^{99}$Tc
\cite{Newman:1976sw,Ullman:1978xy}, i.e.\ $\xi_1\hat{n}_1$. In
contrast, this work concerns $\xi_2\hat{n}_2$. A reevaluation of
the earlier work is given in Ref.~\cite{Jacob} relating it to LIV
in allowed decays~\cite{JN2013}. Within this theory the limits
found from Refs.~\cite{Newman:1976sw,Ullman:1978xy} and our work are complementary: they probe different
combinations of parameters characterizing the LIV interaction.

In the present experiment, we test Lorentz invariance by looking
for a change in the decay rate of allowed $\beta$ decay of
$^{20}$Na when reversing the orientation of the nuclear spin
$\vec{I}$, thus putting a limit on the parameter $\xi_2\hat{n}_2$.
Choosing the polarization direction (denoted by $\hat{z}$) to be perpendicular to the
horizontal plane, a fixed preferred
direction $\hat{N}$ with components \{$N^1, N^2, N^3$\} in the
Sun-centered frame \cite{Kostelecky:2008ts} will lead to an
observation in the laboratory frame as a time dependence of the form
\begin{eqnarray}\label{suncenteredframe}
   \hat{z} \cdot\hat{n}_2(t) &=& N^1\sin\theta\cos\omega_\oplus t +N^2\sin\theta\sin\omega_\oplus t + N^3\cos\theta .
\end{eqnarray}
This has two time-dependent terms with sidereal frequency
$\omega_\oplus$ and one independent of time, with the relative amplitudes
determined by the colatitude of the experiment, $\theta$.

\begin{figure}
  \includegraphics[width=0.6\columnwidth]{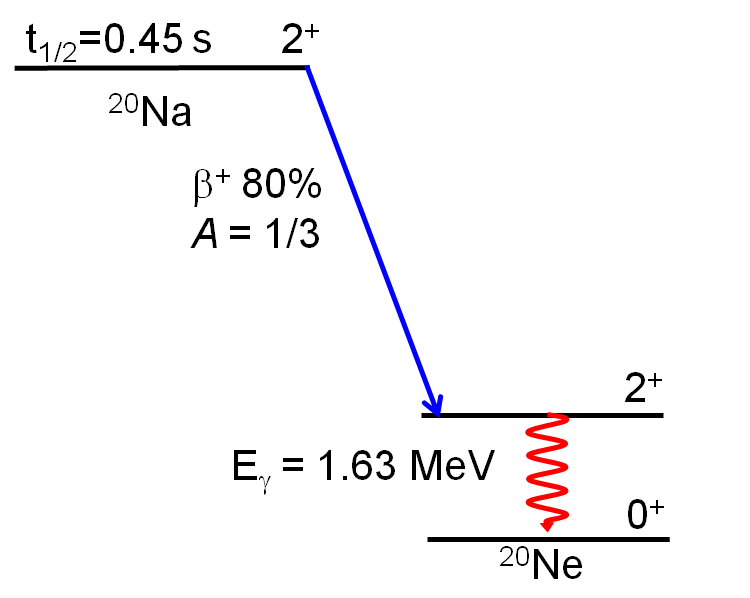}\\
  \caption{Relevant part of the decay scheme of $^{20}$Na.}\label{levelscheme}
\end{figure}
The relevant properties \cite{Tilley} of ${}^{20}$Na decay are shown in
Fig.~\ref{levelscheme}. $^{20}$Na is a short-lived isotope
(t$_{1/2}= 0.45$~s), which decays mostly (80\%) via $\beta^+$
emission in a $2^+ \rightarrow 2^+$ Gamow-Teller transition, for
which $A=1/3$. The endpoint energy for this transition is
11.2 MeV. By measuring the $\beta$ asymmetry we determine the
magnitude of the polarization $\langle\vec{I}\rangle/{I}$. The
decay of the daughter nucleus is observed from the subsequent 1.6
MeV $\gamma$ emission, which can be well discriminated from positron
annihilation radiation at 0.511 MeV. Using the parity-even
$\gamma$ decay, instead of the parity-odd $\beta$ decay, one can
measure the lifetime of the parent
independent of the polarization.
Any residual dependence on polarization, in particular a sidereal dependence, is
 a measure of LIV.
In our analysis we assume that the electromagnetic and strong
interactions obey Lorentz symmetry.

\begin{figure}
\includegraphics[width=\columnwidth]{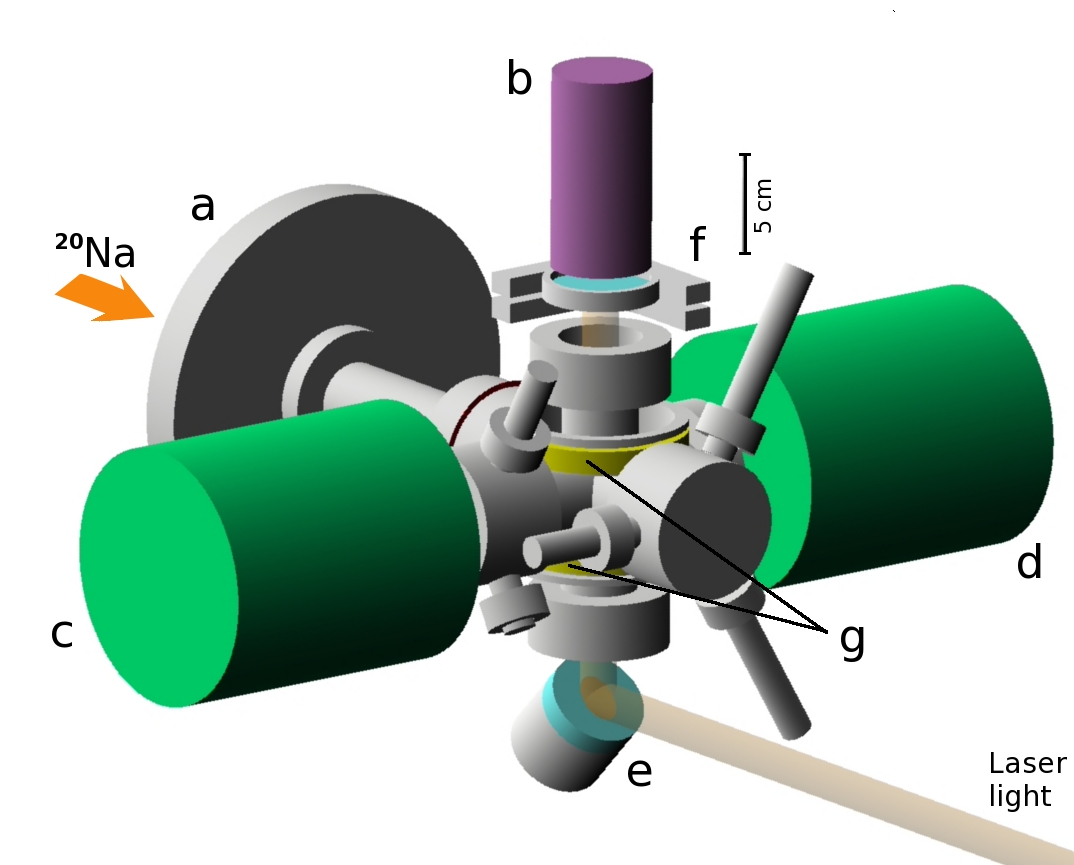}
\caption{Schematic drawing of the experimental setup. Shown are the flange with entrance window (a), 
the $\beta$ detector (b), the NaI $\gamma$ detectors (c,d), optical mirrors (e,f), and coils for the
magnetic holding field (g).}
\label{fig:G4}
\end{figure}
${}^{20}$Na is produced in the ${}^{20}$Ne(p,n)${}^{20}$Na
reaction by colliding a ${}^{20}$Ne beam of 23 MeV/nucleon with
hydrogen in a liquid-nitrogen-cooled gas target. The resulting
isotopes pass through the TRI$\mu$P isotope separator facility to
obtain a clean ${}^{20}$Na beam \cite{Traykov}. The energetic
particles are stopped in a cell with Ne buffer gas at 2 atm. 
With adjustable aluminum degrader foils in the beam line
the beam's stopping distribution is centered in
the cell by maximizing the $\beta$ count rate.

The Na atoms are polarized by optical pumping.
Therefore, the neon buffer gas is cleaned with a liquid-nitrogen-filled cryotrap and a gas purifier cartridge.
A heatable dispenser with natural sodium is mounted inside the buffer-gas
cell. The natural sodium binds residual chemically active
contaminants in the gas and prevents the radioactive sodium
from forming molecules.

The stopped ${}^{20}$Na atoms in the center of the gas cell (see
Fig.~\ref{fig:G4}) are optically pumped into a ``stretched'' state
in which the electronic and nuclear spins are both aligned
vertically by a magnetic holding field of about 2\,mT provided by
two coils. A circularly polarized laser beam with \mbox{589 nm}
wavelength is sent through the buffer gas cell. The pumping can be
achieved both via the $^2{\rm S}_{1/2}- ^2{\rm P}_{1/2}$ (D$_1$)
and the $^2{\rm S}_{1/2}- ^2{\rm P}_{3/2}$ (D$_2$) transitions. In
our pressure domain it was found from simulations that this leads
to polarizations with opposite sign of about 95\% (D$_1$) and
$-75$\% (D$_2$) \cite{Dijck}. The latter transition has been used
in this experiment. Beam blockers are employed to switch the
helicity of the laser beam going through the cell.
The nuclear
polarization $P$ is verified by detecting $\beta^{+}$ particles
emitted in the upward direction with a plastic scintillator. The
$\beta$ detector is set to
trigger on minimum-ionizing particles. The contribution of
$\gamma$ rays to its count rate can be neglected for the
polarization measurement. The $\gamma$-decay rate is measured
with two NaI detectors, 5 inches in diameter. The detector
threshold is set above the annihilation peak at 0.511 MeV to a
level of about 1 MeV where the $\gamma$ spectrum is relatively
flat. The $\gamma$ detectors are positioned perpendicular to the
polarization direction to symmetrize the setup with respect to the
$\beta$ particles.
The $\beta$- and $\gamma$-ray asymmetry are obtained by measuring
the count rates in dead-time-free scalers. The scalers are read
every 1 ms. A fraction of the analog data is digitized for
data-quality control.

The $\beta$ asymmetry is obtained as

\begin{equation}\label{eq:asym}
\mathcal{A}_\beta=PK_A \approx \frac{R_\beta^\uparrow -
R_\beta^\downarrow}{R_\beta^\uparrow + R_\beta^\downarrow},
\end{equation}
where due to the high $\beta$ energies and the placement of the
$\beta$ detector the analyzing power $K_A\approx A=1/3$ and
$R_\beta^{\uparrow(\downarrow)}$ is the count rate for up(down)
polarization.

\begin{figure}
 \centering \includegraphics[width=0.85\columnwidth]{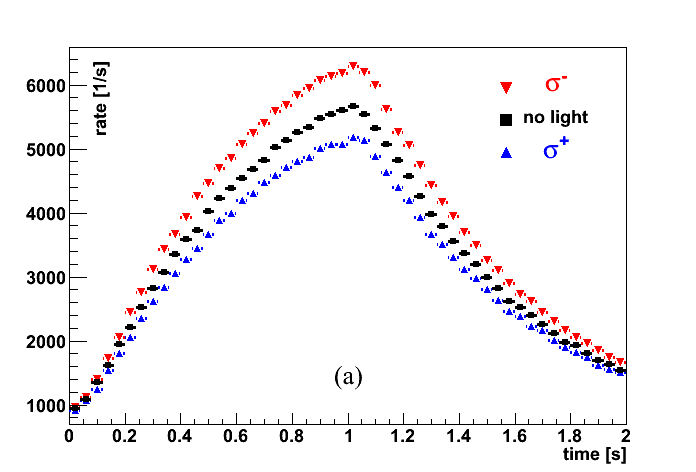}\\
  \includegraphics[width=0.9\columnwidth]{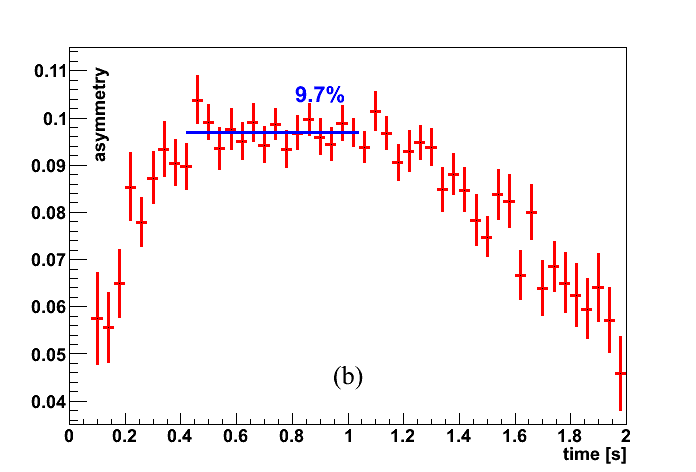}
\caption{(a) Rates detected with the $\beta$-detector for
    different helicity states of the laser light with a pulsed beam of
    ${}^{20}$Na (0 - 1 s ``on'', 1 - 2 s ``off''). Rates have been averaged
    over a data taking period of 35 min. (b) Corresponding asymmetry $\mathcal{A}_\beta$ from
    Eq.~(\ref{eq:asym}). Note the offset vertical scales.}
\label{fig:PS}
\end{figure}

The cyclotron was operated with a cycle of 1 s ``on'' - 1 s ``off,'' and 0.4 s to change polarization.
Data were recorded in sequences of three ``on''-``off'' cycles:
two with opposite laser light helicity and one in which no laser
light entered the buffer-gas cell. In this last period, no
polarization of nuclei is expected. Figure~\ref{fig:PS}a shows the
rates measured by the $\beta$ detector for different helicities of
the laser light. Figure~\ref{fig:PS}b shows the corresponding
$\beta$ asymmetry, $\mathcal{A}_\beta$, defined in
Eq.~(\ref{eq:asym}). Once the beam is on, $\mathcal{A}_\beta$ quickly rises to a
plateau of approximately $10\%$, implying 30\% polarization. If
all atoms are in the laser field a polarization
of about 75\% is expected.
The observed polarization is lower because of diffusion of the
$^{20}$Na atoms out of the laser light and molecule formation of
sodium with residual chemically active contaminants in the buffer
gas. This is consistent with the decreasing $\beta$ asymmetry
when the $^{20}$Na beam is switched off, as seen in
Fig.~\ref{fig:PS}b. When the dispenser with natural Na is heated,
the polarization is also lower, due to convection in the buffer
gas and/or additional unwanted materials that are heated off
surfaces. After turning off the dispenser the polarization in the
beam-``on'' cycles maximizes and then slowly drops in the course
of hours. We also note that not all stopped Na ions neutralize
due to the ionization-energy difference of Na and Ne,
cf.~Ref.~\cite{Dendooven}. For these reasons we make the
simplifying assumption that at the end of the production cycle
30\% of the $^{20}$Na is fully polarized while the remaining
fraction is not.

The $\gamma$-ray rates, $R^{\uparrow\downarrow}_\gamma$, for the two
polarization directions are used to determine the lifetimes $\tau^{\uparrow(\downarrow)}$ from which
the LIV $\gamma$-ray
asymmetry, $\mathcal{A}_\gamma$, is then constructed as
\begin{equation}\label{asymgamma}
\mathcal{A}_\gamma =
\frac{\tau^\uparrow - \tau^\downarrow}{\tau^\uparrow +
\tau^\downarrow}=P\xi_2\hat{n}_2\cdot\hat{z},
\end{equation}
where $\hat{n}_2\cdot\hat{z}$ is given in Eq.\ (\ref{suncenteredframe}).

The $\gamma$-count-rate spectra of the 2-second cycles are combined as in Fig.\ 3a,
accumulated in half-hour bins into three spectra: one for the unpolarized case and one for each of the
two polarization directions. 
A count rate dependence on the gain of the photomultiplier tube could not be avoided, 
making the energy threshold of the trigger count rate dependent.
In addition, pile-up occurs. Both dependencies require correction that is quadratic in the rate.
The corresponding factor is determined using the non-polarized data set.
After this correction, consistent fits with a single effective lifetime were obtained.
The two $\gamma$ detectors have different responses and are analyzed separately;
in the following the data are shown for one detector.

\begin{figure}[t]
\centering
  \includegraphics[width=0.9\columnwidth]{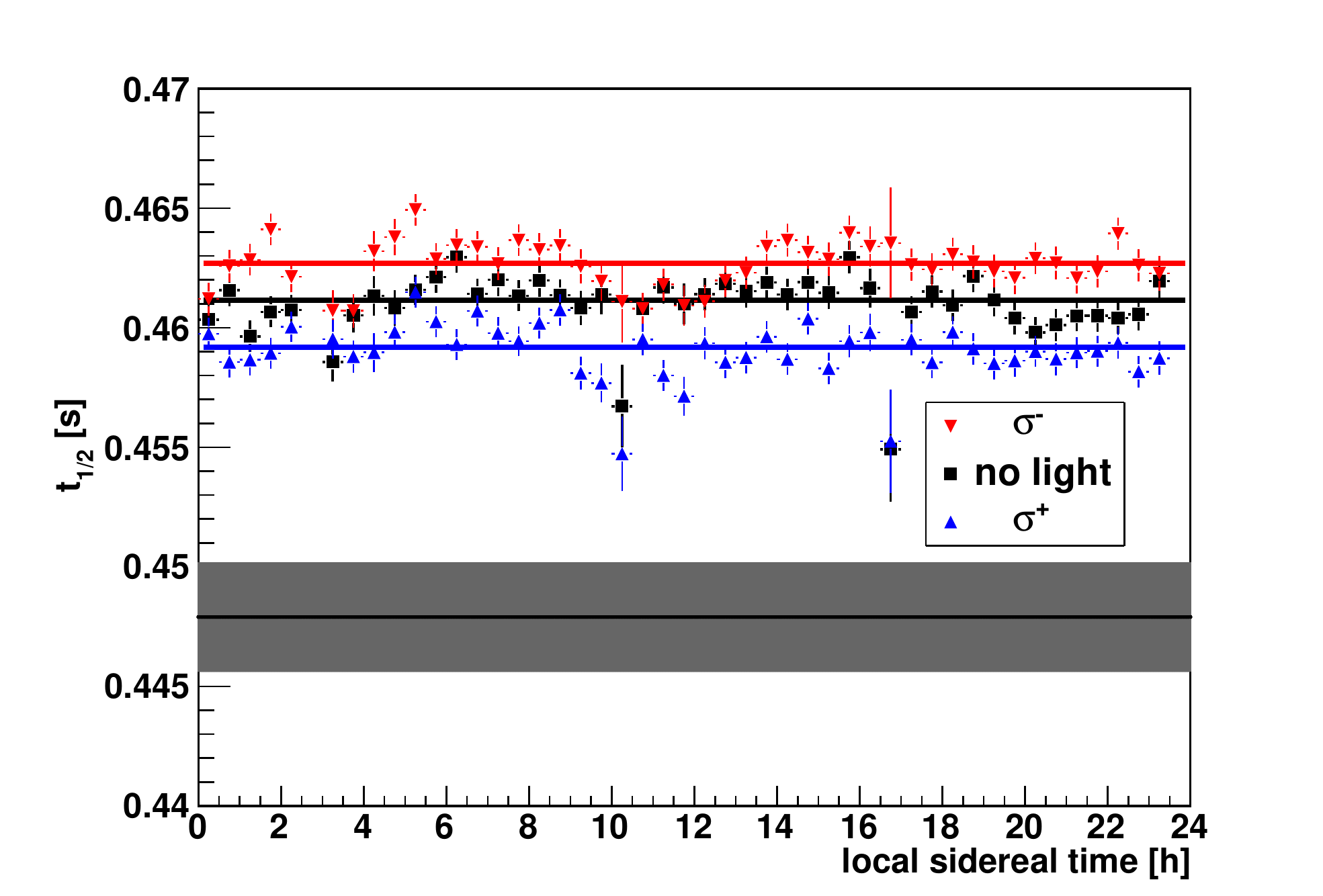}
\caption{Extracted lifetimes as a function of local sidereal time
for the three polarization states (up, off, down). The accepted
value for the lifetime of $^{20}$Na is indicated by the dark band.
} \label{lifetimes}
\end{figure}

A residual dependence on count rate still remained (see Fig.\ \ref{lifetimes}).
The average half-life for the nonpolarized data 
deviates 12 ms from the literature value ($447.9 \pm 2.3$ ms \cite{Tilley}),
while the results between the two polarization conditions differ by 3.1 ms.
Indeed, in Fig.~\ref{lifetimes} we see systematic variations in the fitted lifetimes
occurring for all three modes simultaneously.
This systematic error has serious consequences for the $\gamma$-ray asymmetry.
The $\gamma$ detectors are not blind to positron emission;
a fraction of the $\beta$ particles enter the detectors directly and, in addition,
positrons that annihilate in the surrounding matter deposit radiation.
Although the amount of energy can be too low to cross the threshold,
coincident summing with incompletely detected 1.6 MeV photons produces additional triggers.
Because the polarization is time dependent this results in a changing count rate
that makes the lifetime dependent on polarization.
The effect is dependent on the polarization direction because the polarization axis
is not exactly perpendicular to the axis of the $\gamma$ detectors, 
the detectors have different efficiencies, and the matter surrounding these detectors is not symmetric.
However, the
production and decay cycles are so short that the conditions
within a few cycles are very similar and do not accumulate into
differences on the sidereal time scale.
Dividing the data without polarization in two sets separated by 6 seconds,
an asymmetry constructed from these two sets as in Eq.~(\ref{asymgamma}) shows no sidereal dependence
within the statistical accuracy.
The setup is thus well suited to search for the sidereal dependence.
For limiting the constant term of Eq.~(\ref{suncenteredframe}) it is not suited.

\begin{figure}[t]
\centering
  \includegraphics[width=0.9\columnwidth]{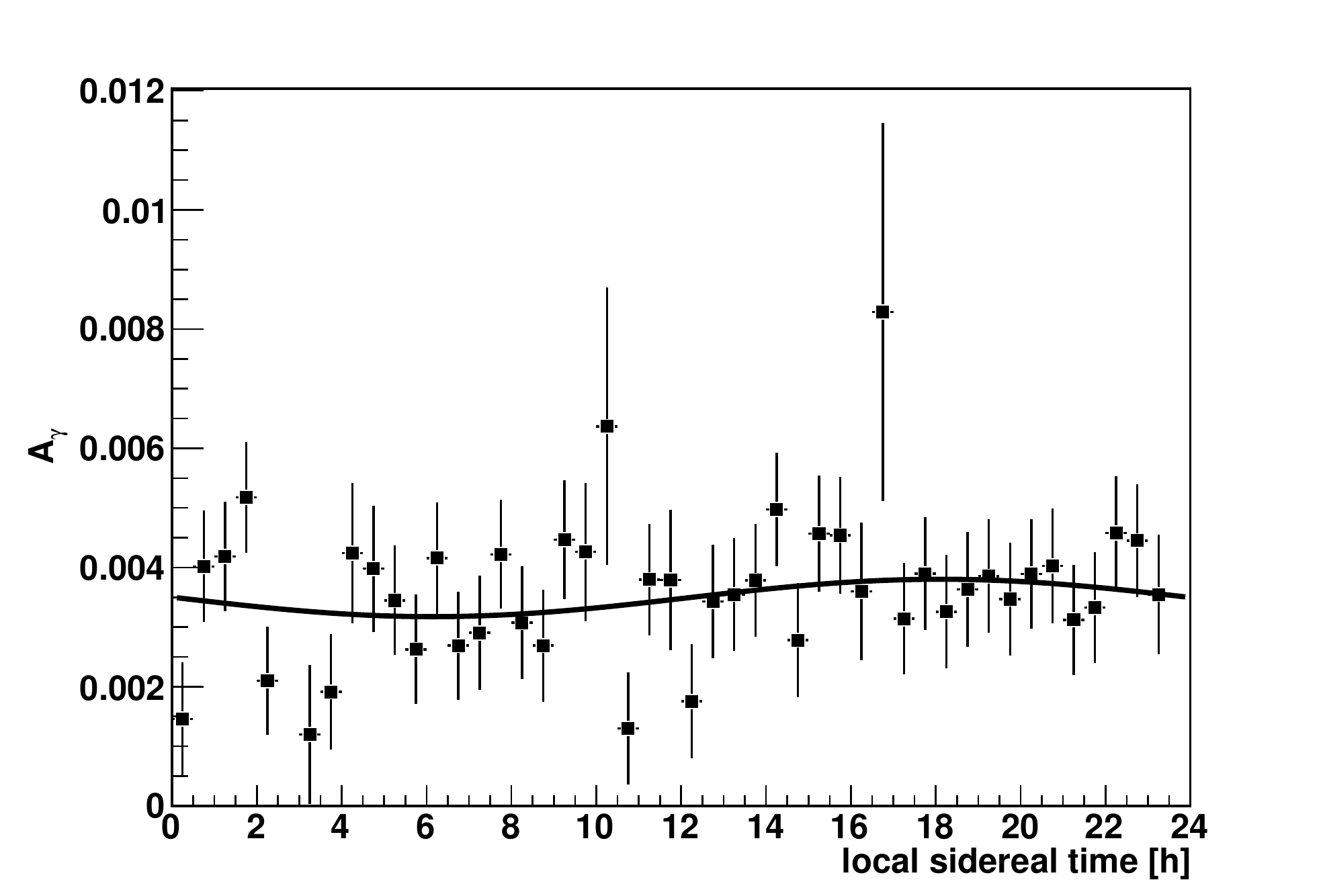}
\caption{The $\gamma$-ray asymmetry determined from the half-life
differences between up and down polarization. The origin of the
graph has been chosen to coincide with local sidereal time zero,
to allow separate limits for the $N^{1,2}$ directions. }
\label{NaI6}
\end{figure}

In Fig.~\ref{NaI6} we show the lifetime asymmetry of
Eq.~(\ref{asymgamma}). The offset in the data has been explained
above. The sidereal time dependence obtained from a fit to the
data does not deviate statistically significant from zero.
Combining the data of both detectors gives a bound of
$|\mathcal{A}_\gamma| < 4 \times 10^{-4}$ at 95\% confidence. The
average polarization is 23\% with an uncertainty due to the
averaging procedure over the polarization lifetime of about 5\%.
The colatitude of our institute is $37^\circ$. Therefore, the
anisotropy in the equatorial plane is bound by $|\xi_2|<3\times
10^{-3}$ at 95\% C.L. In the top half of Table~\ref{tab:xi} the
result is given for the two equatorial directions separately.

In the theoretical work \cite{JN2013} it was shown that a rather
general extension of the standard model is obtained by adding a
LIV tensor $\chi^{\mu\nu}$ to the W-boson propagator.
For a Gamow-Teller transition the value
\begin{equation}\label{therealthing}
\xi_2 \hat{n}^l=A\epsilon^{lmk}\chi_i^{mk}=A\tilde{\chi}^l_i
\end{equation}
was found. Here $A$ is the $\beta$-asymmetry parameter, the index
$i$ refers to the imaginary part of $\chi^{\mu\nu}$, while the
indices $(k,l,m)$ refer to the spacelike components of
$\chi^{\mu\nu}$. Transformed to the Sun-centered frame $\chi^{kl}
\rightarrow X^{kl}$, which can be written in terms of 
standard model extension (SME) parameters \cite{Colladay:1998fq}.
$|\tilde{X}_i^l|$ depends on the antisymmetric part of
$k_{\phi\phi}$ and on $k_{\phi W}$~\cite{JN2013}. The
corresponding expressions and the limits at 95\% C.L. are given in
the bottom half of Table~\ref{tab:xi}. A value for $\tilde{X}_i^3$
cannot be obtained with the present setup. In Ref.\ \cite{JN2013}
it is shown how the real part of $\chi^{\mu\nu}$ can be obtained
in different $\beta$-decay measurements exploiting also the
different dependence of Fermi and Gamow-Teller transitions on
$\chi^{\mu\nu}$.

A setup optimized for $\tilde{X}_i^1$ and $\tilde{X}_i^2$ should have its polarization 
perpendicular to the Earth's rotation axis.
Such a measurement, where, in addition, the $\gamma$ detection
is much less sensitive to $\beta$ particles is currently being analyzed.

In summary, $\beta$ decay can be exploited to search for Lorentz invariance violation in the weak interaction,
which has not been searched for yet in any systematic way.
The framework developed in Ref.~\cite{JN2013} makes this possible by parametrizing the LIV
in terms of a background tensor $\chi^{\mu\nu}$.
The current work is the first attempt to put limits on the magnitude of $\chi^{\mu\nu}$.
A limit was set on the imaginary part of the spacelike component of $\chi^{\mu\nu}$
by searching for a variation in the decay rate of $^{20}$Na nuclei as function of their spin direction.
A limit of a sidereal dependence in the equatorial plane of
$3\times 10^{-3}$ at 95\%~C.L. was found, independent of theory.
The theoretical framework implies that with this experiment and a
variety of other experiments one can limit the magnitude of the
LIV tensor further.

\begin{table}
  \begin{center}
    \caption{Limits on $\xi_2N^{1,2}$ and the corresponding SME parameters in the Sun-centered frame.}\label{tab:xi}
    \begin{tabular}{|c| r| r|}
      \hline
      \hline
      Coefficient & \multicolumn{1}{c|}{Value} & \multicolumn{1}{c|}{95\% C.L.\ interval} \\
      \hline
      $\xi_2N^{1}$ & $(-12\pm9) \times 10^{-4}$ & $[-29,+6]\times 10^{-4}$ \\
      $\xi_2N^{2}$ & $( -3\pm8) \times 10^{-4}$ & $[-19,+14]\times 10^{-4}$\\
      \hline
      $\tilde{X}^1_i$
      &$2(k^A_{\phi \phi})^{32}+\frac{1}{g}(k_{\phi W})^{32}$
      &$[-9,+2]\times
      10^{-3}$\\
      $\tilde{X}^2_i$
      &$2(k^A_{\phi\phi})^{13}+\frac{1}{g}(k_{\phi W})^{13}$&$[-6,+4]\times
      10^{-3}$\\
      \hline
      \hline
    \end{tabular}
  \end{center}
\end{table}

We thank the KVI cyclotron staff for providing the beam and L.\
Huisman for technical support. This research was financially
supported by the ``Stichting voor Fundamenteel Onderzoek der
Materie (FOM)'' under Program 114 (TRI$\mu$P) and ``FOM
projectruimte'' 08PR2636-1.

\end{document}